\documentclass[twocolumn,showpacs,prl,amsmath,amssymb]{revtex4}
\usepackage{graphicx}% Include figure files
\usepackage{bm}
\begin{document}

\title{Quantum Phases of the Extended Bose-Hubbard Hamiltonian:  
Possibility of a Supersolid State of Cold Atoms in Optical Lattices}
\author{V.W. Scarola and S. Das Sarma}
\affiliation{Condensed Matter Theory Center, 
Department of Physics, University of Maryland,
College Park, MD 20742-4111}

\begin{abstract}
Cold atom optical lattices typically simulate zero-range Hubbard 
models.  We discuss the theoretical possibility of using 
excited states of optical lattices to generate {\em extended} range 
Hubbard models.  We find that bosons confined to higher 
bands of optical lattices allow for a rich phase diagram, including
the supersolid phase.  Using Gutzwiller, mean field theory we establish the 
parameter regime necessary to maintain metastable states generated  
by an extended Bose-Hubbard model.
% of cold bosons
%confined to higher bands of optical lattices.   
\end{abstract}
\maketitle

Bose condensed cold atom systems in optical 
lattices \cite{Phillips} are increasingly
serving as beautiful (and practical) laboratories for studying quantum
phases and quantum phase transitions in strongly correlated model
Hamiltonians of great intrinsic interest.  In particular, the very low
temperature; the absence of disorder, dirt and defects; and
(essentially) complete control over the system parameters (and
therefore the applicable Hamiltonian) combine to make cold atoms
in optical lattices an ideal system to experimentally test the 
predictions of various interacting quantum Hamiltonians which
originated as model (and often rather unrealistic) descriptions for
condensed matter physics problems.  For example, the Mott insulating
phase and the superfluid phase of a Bose Hubbard model have been 
demonstrated in the cold atom optical lattices \cite{Jaksch,Greiner}.  

In spite of the impressive success of the cold atom systems in studying
the quantum phases of strongly correlated Hamiltonians, there has been
one important limitation.  Cold atoms in optical lattices usually
represent essentially zero-range systems where the correlation 
(or, equivalently, the inter-particle interaction) is effectively
on-site only, being parameterized by a single interaction
energy $U$ (the so-called Hubbard $U$), so that the system Hamiltonian
is the Hubbard model characterized by a single dimensionless
coupling parameter $t/U$, where $t$ is the quantum tunneling or
hopping strength.  In this letter we propose a specific and practical
scheme to generalize the cold atom Hamiltonian to an {\em extended}
Hubbard model, where both on-site ($U$) and longer-range ($V$)
inter-particle interactions compete with the kinetic energy giving
rise to a rich quantum phase diagram which should be experimentally
accessible.  We focus, in particular, on bosonic systems though the
method described here is applicable to spinful, fermionic systems as well.  In 
the bosonic case, our proposed scheme may lead to density 
wave and supersolid quantum phases in addition to the ``usual''
Mott insulating and superfluid phases.  The key idea in our work, enabling
the realization of an extended Hubbard atomic system, is that one
could, by utilizing proper laser excitations of individual cold atom
states in the confining potential \cite{Wilkinson,Phillips2}, use, in 
principle, the {\em excited} confined states (rather than the
lowest level in each individual optical lattice potential minimum) to
form the interacting system.  Such a system would have a natural
extended Hubbard description rather than an on-site description.  We
theoretically obtain the quantum phase diagram of such a system and
predict the exciting possibility of coexisting density
wave and superfluid order, i.e. the supersolid
quantum phase, in the $U-V$ parameter space of a realistic,
extended Bose-Hubbard model.  Our proposed system should
also have important relevance to topological quantum computation 
\cite{Freedman} in optical lattices which has been shown to
be feasible with cold atom lattices provided an extended range
interaction ($V\neq0$) applies.  

Cold atomic gases confined to optical lattices 
offer the unique opportunity to directly probe novel states of matter,
including the supersolid.
In comparison, experimental evidence for supersolid order
in $\text{He}^4$ now exists \cite{Chan}, though conclusive 
identification using current experimental techniques remains elusive 
\cite{Leggett}.  In bosonic cold atom optical lattices, coherence peaks 
in multiple matter wave interference patterns \cite{Greiner2} 
at {\em half} the reciprocal lattice vector 
would provide strong evidence for supersolid order.  Recent 
proposals suggest that nearby Mott (and therefore density wave) 
order may also be directly observed, but through structure in 
noise correlations \cite{Demler} or through Bragg spectroscopy \cite{Clark}.  

We begin with the following second quantized Hamiltonian describing 
bosons in an optical lattice, interacting through a contact interaction:
\begin{eqnarray}
H=\int d^{3}\textbf{r}\Psi^{\dagger}(\textbf{r})
\left[H_0+V_{\text{conf}}
+\frac{g}{2}\Psi^{\dagger}(\textbf{r})\Psi(\textbf{r})\right]\Psi(\textbf{r}),
\label{H}
\end{eqnarray}
where $g=4\pi\hbar^2a_S/m$ is the three dimensional interaction
strength between bosons of mass $m$ and scattering length $a_S$.  The 
single particle part of the Hamiltonian defines the motional 
degrees of freedom through:
$H_0= -\frac{\hbar ^2}{2m}\nabla^2
+V_{\text{lat}}^d({\textbf r}),$
and the confinement potential:
$V_{\text{conf}}=m[\omega_{1}^2(x^2+y^2)+\omega_2^2
z^2 ]/2.$  
$V_{\text{conf}}$ defines the dimension, $d$, of the system.  For 
$d=1$ or $2$ we have $\omega_{3-d}\ll\omega_{d}$.  The 
optical lattice potential is $V_\text{lat}^1(z)
=V_L\left[1-\text{cos}\left(2\pi
  z/a\right)\right]/2$, for $d=1$.  Here the lattice constant 
is $a=\lambda/2$, where $\lambda$ is the wavelength of the laser 
defining the lattice. 
We also consider a square lattice for $d=2$: 
$V_{\text{lat}}^2(x,y)=V_\text{lat}^1(x)+V_\text{lat}^1(y)$.  With
these single particle potentials, the noninteracting problem
separates.  In the direction of strong confinement we, as a first
approximation, assume the harmonic oscillator ground state,  
separated from higher energy levels by 
$\hbar \omega_d$ thereby establishing 
a $d$-dimensional problem in the remaining
coordinates.  The dimensionless interaction strength becomes 
$\bar{g}_d=(g/E_R)(m\omega_d /h)^{\frac{3-d}{2}}(\pi /a)^d$.
The over-bar indicates dimensionless units: $\bar{\bm{r}}=\pi \bm{r}/a$
and $\bar{H_0}=H_0/E_R$, where $E_R=h^2/ 2m\lambda^2$.  

\begin{figure}
\includegraphics[width=2.5in]{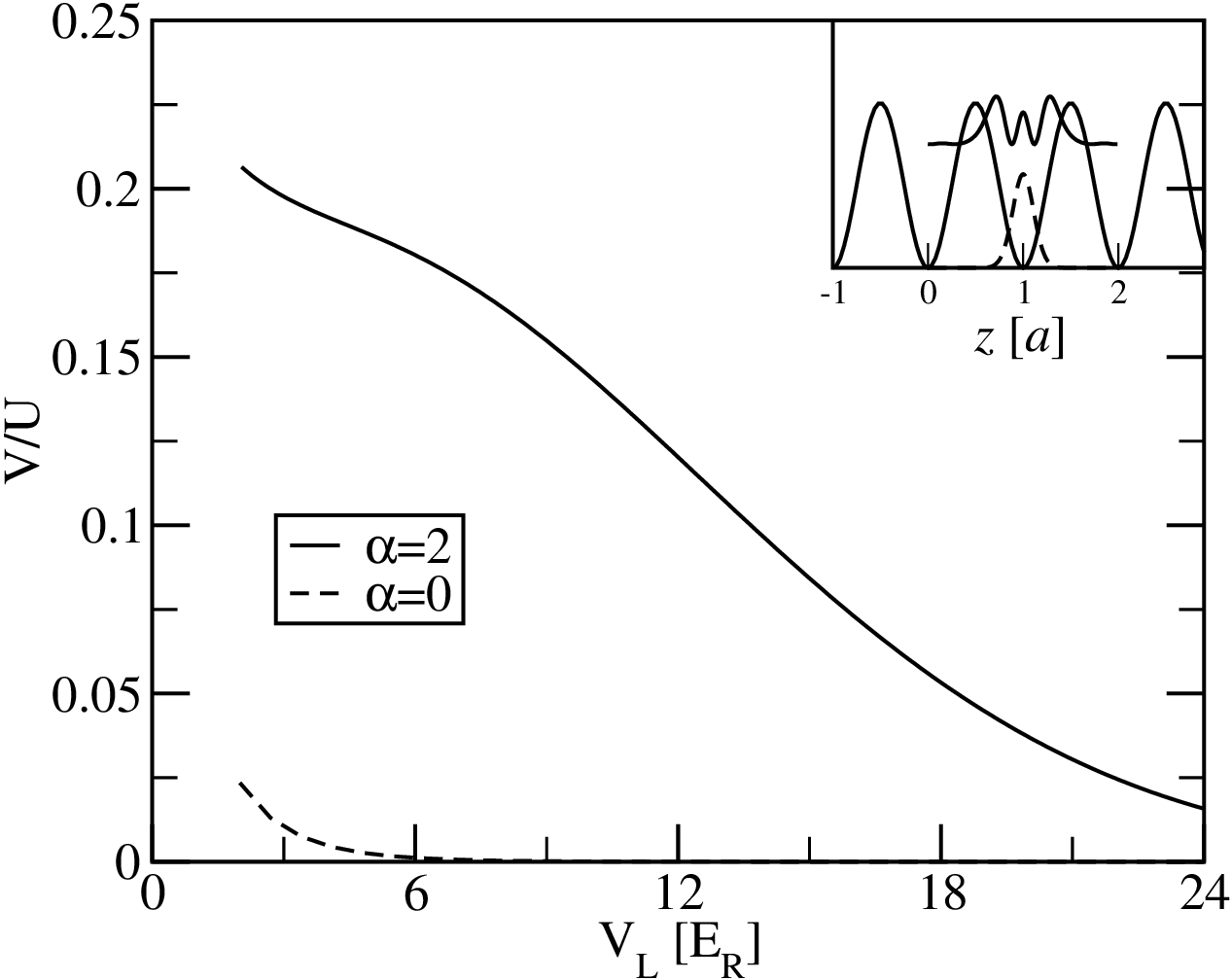}
\caption{ 
The ratio of interaction matrix elements versus lattice depth in
the lowest (dotted line) and the third (solid line) one dimensional 
bands.  The inset shows the square of the Wannier
functions in the lowest (dotted line) and the third (solid line) 
bands plotted as a function of distance against
a confining, sinusoidal lattice.  
\label{VoU}}
\end{figure}

Along the directions of weak confinement the noninteracting problem 
defines a Bloch equation (excluding the
confinement potential).
The exact solutions, $\Phi_{\bm{k},\bm{\alpha}}$,
can be written in terms 
of Mathieu functions with wavevector $\bm{k}$ in band
$\bm{\alpha}$ \cite{Slater}.  From the Bloch functions we 
define the Wannier functions:
$
w_{i,\bm{\alpha}}
=N_s^{-1/2}\sum_{\bm{k}}\exp{(-i\bm{k}\cdot\bm{\delta}_i)}
\bar{\Phi}_{\bm{k},\bm{\alpha}}(\bar{\bm{r}}), 
$
for $N_s$ sites at locations $\bm{\delta}_i$.  
The 
Wannier functions 
localize in 
the ``atomic'' limit for large lattice depths, 
$(V_L/E_R)^\frac{1}{4}\rightarrow\infty$, where the bands reduce  
to the harmonic oscillator energy levels.  In a 
band, energetically, near the lattice 
maximum the density of two Wannier
functions in neighboring sites can have strong overlap.  The inset 
of Fig.~\ref{VoU} depicts two situations, showing
the square of the Wannier functions for $d=1$ in the bands
$\alpha=0$ (dotted line) and $\alpha=2$ (solid line), plotted as a 
function of distance against a host lattice with height $V_L=20E_R$.
The large overlap between nearest neighbor basis states suggests that atoms 
confined to higher bands offer the unique possibility of generating 
extended range interactions from an underlying, short-range
interaction.  In what follows we apply 
this technique to construct an extended, bosonic lattice model in the 
Wannier basis.  A recent proposal \cite{Goral} suggests that 
extended range, Bose-Hubbard models may, alternatively, be generated 
with condensates of dipolar bosons in optical lattices.

We now expand the operators in Eq.~(\ref{H}) via:  
$
\Psi=\sum_{i,\bm{\alpha}}
w_{i,\bm{\alpha}}b_{i,\bm{\alpha}}, 
$
where $b_{i,\bm{\alpha}}$ annihilates a boson at site $i$ in
band $\bm{\alpha}$.  The Hamiltonian becomes:
$
H_w=\sum_{\bm{\alpha}}H^{\bm{\alpha}}
+\sum_{\bm{\alpha}\neq\bm{\alpha}'}H^{\bm{\alpha},\bm{\alpha}'}.
$
We first focus on the largest intra-band terms:
\begin{eqnarray}
H^{\bm{\alpha}}= 
-\sum_{<i,j>}
t^{\bm{\alpha}}_{i,j}
\left(b^{\dagger}_{i,\bm{\alpha}}b^{\vphantom{\dagger}}_{j,\bm{\alpha}}
+h.c.\right)
-\mu^{\bm{\alpha}}\sum_i n_{i,\bm{\alpha}}
\nonumber
\\
+U^{\bm{\alpha},\bm{\alpha}}\sum_i n_{i,\bm{\alpha}}(n_{i,\bm{\alpha}}-1)
+4\sum_{<i,j>}V^{\bm{\alpha},\bm{\alpha}}_{i,j}
n_{i,\bm{\alpha}}n_{j,\bm{\alpha}},
\label{Hex}
\end{eqnarray}
where the chemical potential is defined to be 
$
\mu^{\bm{\alpha}}=\mu_0-\langle w_{i,\bm{\alpha}} \vert 
H_0 \vert
w_{i,\bm{\alpha}} \rangle,
$
with $\mu_0$ a constant dependent on confinement.  The number operator
is given by: 
$n_{i,\bm{\alpha}}\equiv 
b^{\dagger}_{i,\bm{\alpha}}b^{\vphantom{\dagger}}_{i,\bm{\alpha}}$.  
The hopping
between nearest neighbors, denoted $<i,j>$, is 
only intra-band and non-diagonal for cubic lattices:
$
t^{\bm{\alpha}}_{i,j}
=-\langle w_{i,\bm{\alpha}} \vert H_0 \vert
w_{j,\bm{\alpha}} \rangle
$.  It is in principle renormalized by the interaction to include 
conditional hopping:
$
t^{\bm{\alpha}}_{i,j}
\rightarrow t^{\bm{\alpha}}_{i,j}
-2{\cal M}_{j,i,i,i}
^{\bm{\alpha},\bm{\alpha},\bm{\alpha},\bm{\alpha}}
(n_{i,\bm{\alpha}}+n_{j,\bm{\alpha}}-1),
$
where 
$
{\cal M}^{\bm{\alpha}_1,\bm{\alpha}_2,\bm{\alpha}_3,\bm{\alpha}_4}
_{i_1,i_2,i_3,i_4}
\equiv(\bar{g}_d E_R/2)\langle
w_{i_1,\bm{\alpha}_1};w_{i_2,\bm{\alpha}_2}
\vert w_{i_3,\bm{\alpha}_3};w_{i_4,\bm{\alpha}_4} \rangle
$
are the interaction matrix elements.  In our study, we concentrate on the
low density regime, $\rho \lesssim 1$, where $\rho$ is the average
number of particles per site.  In this regime we find the conditional
hopping to not change our results significantly.  Along these lines we
have, in Eq.~(\ref{Hex}), omitted double occupancy terms of the form 
$ 
b^{\dagger}_{j,\bm{\alpha}}
b^{\dagger}_{j,\bm{\alpha}}
b^{\vphantom{\dagger}}_{i,\bm{\alpha}}
b^{\vphantom{\dagger}}_{i,\bm{\alpha}}
$
which, as we have also checked, do not contribute significantly at
low densities.  The remaining two terms in Eq.~(\ref{Hex}) 
define the largest contributions to the 
interaction through the on-site, 
$
U^{\bm{\alpha},\bm{\alpha}'}\equiv{\cal M}_{i,i,i,i}
^{\bm{\alpha},\bm{\alpha}',\bm{\alpha},\bm{\alpha}'}, 
$
and nearest neighbor,  
$
V^{\bm{\alpha},\bm{\alpha}'}_{i,j}
\equiv{\cal M}_{i,j,i,j}
^{\bm{\alpha},\bm{\alpha}',\bm{\alpha},\bm{\alpha}'},
$
coefficients.  
%In the lowest band the latter term is much smaller than the 
%former.  But in higher bands the quasi-localized Wannier states
%near the top of the lattice yield a sizable nearest neighbor
%interaction.  
Fig.~\ref{VoU} plots the ratio 
$V^{\bm{\alpha},\bm{\alpha}}_{i,i+1}/U^{\bm{\alpha},\bm{\alpha}}$ 
as a function of the lattice depth for the 
lowest (dotted line) and the third (solid
line) band in one dimension.  In $d=2$ the result remains
the same as long as we compare the $(0,0)$ and $(2,2)$ bands.  From 
Fig.~\ref{VoU} we clearly see that the ratio can be sizable.  We must
therefore incorporate extended Hubbard terms into any lattice model
characterizing particles in higher bands {\em not} in the atomic limit.

We now discuss a four stage gedanken experiment designed to place 
bosons in a higher band of the optical lattice.  The 
prescription we provide here is not unique but serves to minimize band
mixing.  We first consider a partially filled lowest band 
in the atomic limit and with weak interaction
strength, $\bar{g}\ll 1$.  As we know, from the mean field phase
diagram \cite{Fisher} of the Bose-Hubbard 
model ($V=0$ in Eq.~(\ref{Hex})), bosons, in this 
limit, form a superfluid at all $t/U$.

The second step consists of adiabatically 
loading \cite{Phillips2} the atoms
into a higher band, {\em e.g.} $\bm{\alpha}_p=(2,2)$ in $d=2$, by oscillating
the lattice depth at a frequency matching the inter-band energy
difference.  We assume that a large majority of the atoms can be
transferred from the lowest band to a single, higher band. Once 
loaded into a higher band we note that, in our model, there
is no inter-band coupling for a translationally invariant,
non-interacting system in the steady state.  

In the third stage the lattice depth lowers, away from 
the atomic limit, to a point where the Wannier functions 
have some extension into the barriers 
between sites, $V_L\approx 19 E_R$ in Fig.~\ref{VoU}.  This process may be
considered adiabatic if the time scale associated with lowering the
lattice depth of a noninteracting system is much longer than 
$
h/\vert
\mu^{\bm{\alpha}}-\mu^{\bm{\alpha}_p}\vert,
$
where $\bm{\alpha}$ indicates the nearest band.  
We find 
$
\vert \mu^{\bm{\alpha}}-\mu^{(2,2)}\vert
$
to cross zero linearly as a function of $V_L$ 
near $V_L\approx 15.8 E_R$ and $V_L\approx 17.5 E_R$
for $\bm{\alpha}=(3,1)$ and $(4,0)$, respectively.  

In the last stage we increase $\bar{g}_d$.  Recent studies 
\cite{Jaksch,Greiner,Paredes} have, quite differently, reached the strongly
correlated regime, in the lowest band, by tuning the ratio
$t/U$ with $V_L$.  We, however, require the lattice depth to remain in a
narrow regime.  We assume that the interaction strength itself can
be tuned through, for example, a Feshbach resonance.  In what follows
we study Eq.~(\ref{Hex}), in the range $\bar{g}_d\sim 100-0$.  We 
then analyze inter-band effects induced by large interaction strengths.  

We consider several possible ground states of Eq.~(\ref{Hex}) and
focus on the two dimensional square lattice.  The ground states of this
model contain four types of order, in the absence of disorder 
and at zero temperature:  Superfluid order $<b_i>$, Mott order $<n_i>$,
checkerboard density wave order 
$                
(-1)^{(\delta^x_i+\delta^y_i)}\left[ <n_i>-\rho \right],
$
and supersolid order, where superfluid and density wave 
order coexist.  
Nonzero superfluid order arises from a spontaneously
broken gauge symmetry.  We 
note that the host lattice corrugates the superfluid density at wavevectors  
corresponding to the reciprocal lattice vector.
When phase fluctuations become strong 
Mott order persists at integer densities.  

The extended interaction term frustrates the Mott and superfluid
phases leading to spontaneous translational symmetry breaking, {\em e.g.}  
the $\rho=1/2$ density wave phase, ordered 
at half the reciprocal lattice vector. 
%It is important
%to note that the density wave discussed here 
%is a ``true'' solid in that peaks in the structure factor 
%of this phase arise from spontaneously broken, translational 
%symmetry.  
In contrast, a deep host lattice 
induces the corrugation in the Mott and superfluid phases.  The 
fourth phase, the supersolid, arises from dual 
spontaneous symmetry breaking (both translational and gauge symmetry) 
inherent in coexisting density wave and superfluid order.  
%One may compare 
%the supersolid phase to a state associated with macroscopic 
%occupation of a lattice mode with similar quantum 
%numbers but with no spontaneously broken translational 
%symmetry:  the $k=\pm\pi/a$ minima of the $(1,1)$ band, for example.  

\begin{figure}
\includegraphics[width=2.5in]{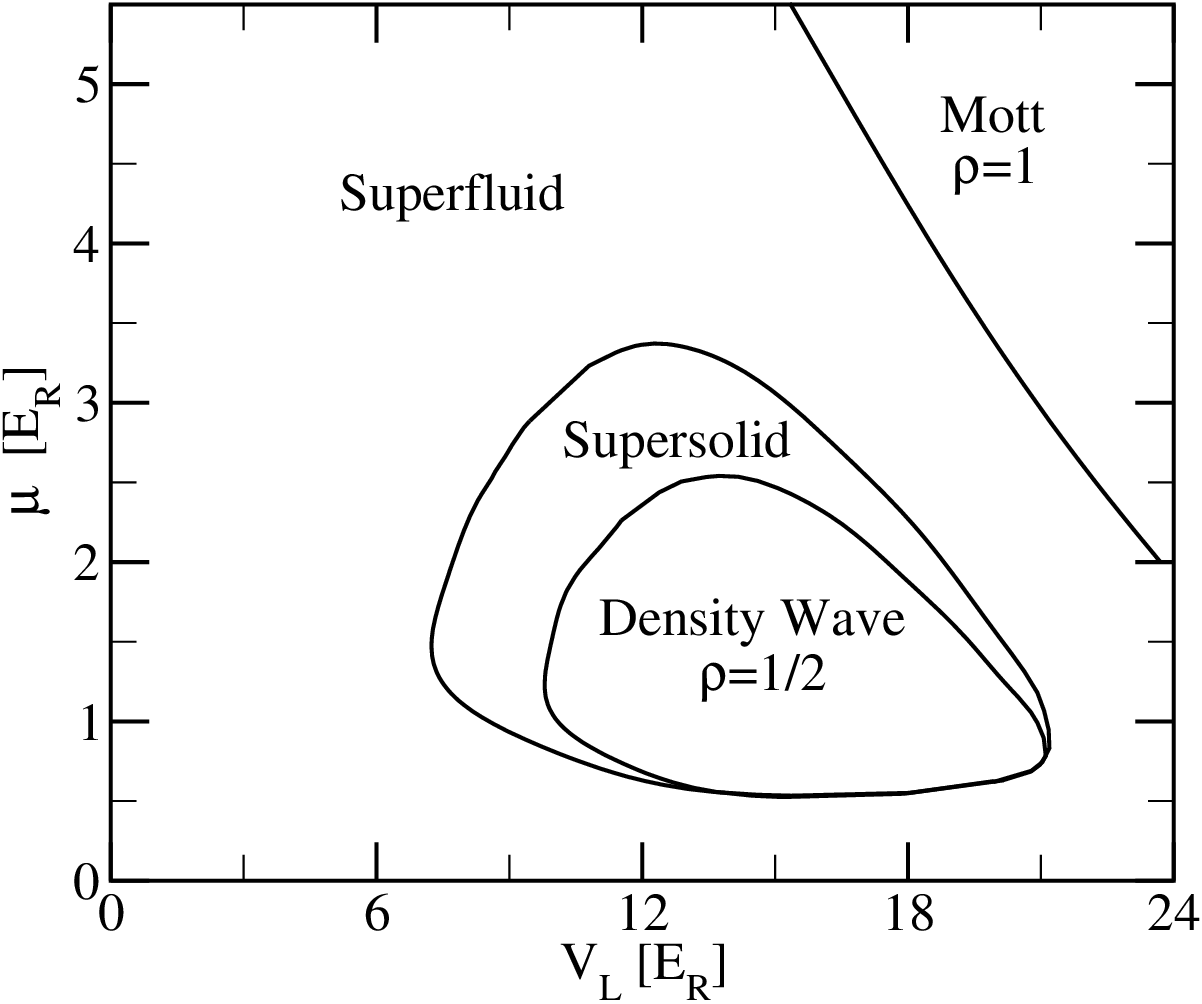}
\caption{ 
Zero temperature, mean field phase diagram of the extended Bose-Hubbard
model, Eq.~(\ref{Hex}), as determined by the band structure 
of a two-dimensional, square, lattice in the $(2,2)$ 
band with interaction strength $\bar{g}_2=50$.  
%The chemical potential is plotted 
%along the $y$-axis and the 
%lattice depth is plotted along the $x$-axis, both in units of photon
%recoils, $E_R$.
%The hopping {\em decreases} along the $x$-axis mirroring the usual,
%hopping dependent Bose-Hubbard phase diagram.  
\label{phase}}
\end{figure}

We now discuss our solution of Eq.~\ref{Hex} in the band
$\bm{\alpha}_p=(2,2)$.  We solve $H_0$ exactly to obtain 
the matrix elements.  We use a Gutzwiller variational
ansatz \cite{Rokshar,Jaksch} equivalent to a mean field decoupling
of $H^{\bm{\alpha}_p}$:
$
\psi_{\bm{\alpha}}
=\prod_{i}\left[\sum_{N_{{i,\bm{\alpha}}}=0}^{\infty}
f_{N_{{i,\bm{\alpha}}}}\vert N_{{i,\bm{\alpha}}} \rangle\right],
$
where the variational parameters, $f_{N_{{i,\bm{\alpha}}}}$, may vary
over distinct sublattices and weight Fock states with
$N_{{i,\bm{\alpha}}}$ particles.
We minimize  Eq.~(\ref{Hex}) with respect
to $f_{N_{{i,\bm{\alpha}}}}$ keeping 
enough $N_{{i,\bm{\alpha}}}$ to ensure convergence of the total energy.  Note 
that $t_{i,j},\mu,V_{i,j},$ and $U$ depend only on 
$V_L/E_R, \bar{g}_d,$ and $\mu_0/E_R$.  Fig.~\ref{phase} shows the two
dimensional phase diagram for $\rho\lesssim 1$ in the principal band, 
$(2,2)$, with $\bar{g}_2=50$.
The $y$-axis plots the chemical potential and the $x$-axis the lattice
depth, both in units of $E_R$.  The hopping and the extended Hubbard
coefficients decrease with increasing lattice 
depth.  Accordingly, we find Mott order at large
lattice depths.  
%For intermediate lattice depths
%the extended Hubbard term competes with other terms.  
Supersolid 
and density wave order appear for $\bar{g}_2\gtrsim 40$.  
%We emphasize that for
%$\rho\lesssim 1$ the conditional hopping terms and terms of the form 
%$ 
%b^{\dagger}_{j,\bm{\alpha}}
%b^{\dagger}_{j,\bm{\alpha}}
%b^{\vphantom{\dagger}}_{i,\bm{\alpha}}
%b^{\vphantom{\dagger}}_{i,\bm{\alpha}},
%$
%have been included in generating the phase diagram but play 
%an insignificant role and may be excluded.  
The supersolid phase appears upon doping of the 
density wave phase at $\rho=1/2$ and not the Mott 
phase, consistent with the results of 
Ref.~\cite{Prokofev}.  We add that in one
dimension the phase diagram is nearly identical for the same set of
parameters.  However, it is by now well established \cite{White} that
fluctuations destroy supersolid order in one dimension.  In both 
one and two dimensions (though
more so in two dimensions) nearby bands energetically
approach the principal band at low lattice depths.

We now study interaction induced, inter-band effects.  Our single band
approximation, Eq.~(\ref{Hex}), comes into question as we lower the lattice depth.  We
study, for $d=2$, mixing with the two nearest bands $(3,1)$ and
$(1,3)$.  In principle, mixing with nearby bands can alter the phase
diagram.  However, if only a small fraction of the atoms occupy
neighboring bands we may then safely assume that the phase
diagram remains qualitatively the same. 
We ask whether or not the ground states of Fig.~\ref{phase} in
the principal band $\bm{\alpha}_p=(2,2)$ favor scattering 
processes coupling neighboring bands \cite{Rolston}.  The dominant 
inter-band terms are: 
\begin{eqnarray}
&&H^{\bm{\alpha},\bm{\alpha}'}=4U^{\bm{\alpha},\bm{\alpha}'}\sum_{i} 
n_{i,\bm{\alpha}}n_{i,\bm{\alpha}'}
+4\sum_{<i,j>}
V^{\bm{\alpha},\bm{\alpha}'}_{i,j}
n_{i,\bm{\alpha}}n_{j,\bm{\alpha}'}
\nonumber
\\
&&+\sum_{i}{\cal M}_{i,i,i,i}
^{\bm{\alpha}'\pm\bm{\Delta},\bm{\alpha}',\bm{\alpha},\bm{\alpha}}
\left(b^{\dagger}_{i,\bm{\alpha}'\pm\bm{\Delta}}
b^{\dagger}_{i,\bm{\alpha}'}
b^{\vphantom{\dagger}}_{i,\bm{\alpha}}
b^{\vphantom{\dagger}}_{i,\bm{\alpha}}
+h.c.\right),
\label{Hmix}
\end{eqnarray}
where the matrix element in the last term ensures conservation of band
index (arising from conservation of lattice 
momentum).  We have $\bm{\Delta}=(-2,2)$ for 
$
\bm{\alpha},\bm{\alpha}'\in \{(2,2),(1,3),(3,1)\}
$
in two dimensions.   

The last term in Eq.~(\ref{Hmix}) takes two particles from the same
site in the principal band and ``scatters'' them to neighboring bands 
and vice versa, when applied to a state initially in ${\bm{\alpha}_p}$.
It acts as the dominant inter-band scattering 
mechanism \cite{inter}.  We calculate the probability 
of such an event through first order perturbation
theory.  Consider two states $\psi_{\bm{\alpha}_p}$ and:   
$\psi_e\equiv{\cal N}^{-1}(b^{\dagger}_{i,\bm{\alpha}_p+\bm{\Delta}/2}
b^{\dagger}_{i,\bm{\alpha}_p-\bm{\Delta}/2}
b^{\vphantom{\dagger}}_{i,\vphantom{\bm{\Delta}/2}\bm{\alpha}_p}
b^{\vphantom{\dagger}}_{i,\vphantom{\bm{\Delta}/2}\bm{\alpha}_p})
\psi_{\bm{\alpha}_p}, 
$ 
where $\cal N$ is a normalization constant.  In the absence 
of dissipation, the probability that two 
particles at any one site occupy neighboring bands 
oscillates in time, $T$:
$
P(T)
=2 \vert \langle \psi_e \vert H_w \vert \psi_{\bm{\alpha}_p} 
\rangle\vert^2
\left[1-\text{cos}\left(E T/\hbar\right)\right]/E^2,  
$
where
$
E\equiv\langle \psi_e \vert H_w \vert \psi_e \rangle
-\langle \psi_ {\bm{\alpha}_p}\vert H_w \vert \psi_{\bm{\alpha}_p}
\rangle.
$
We argue that if the probability remains small, then 
band mixing will be suppressed.
Note that, with a large interaction strength, $E$ is {\em not} 
equal to the single-particle, self energy difference 
between bands.  

\begin{figure}
\includegraphics[width=2.4in]{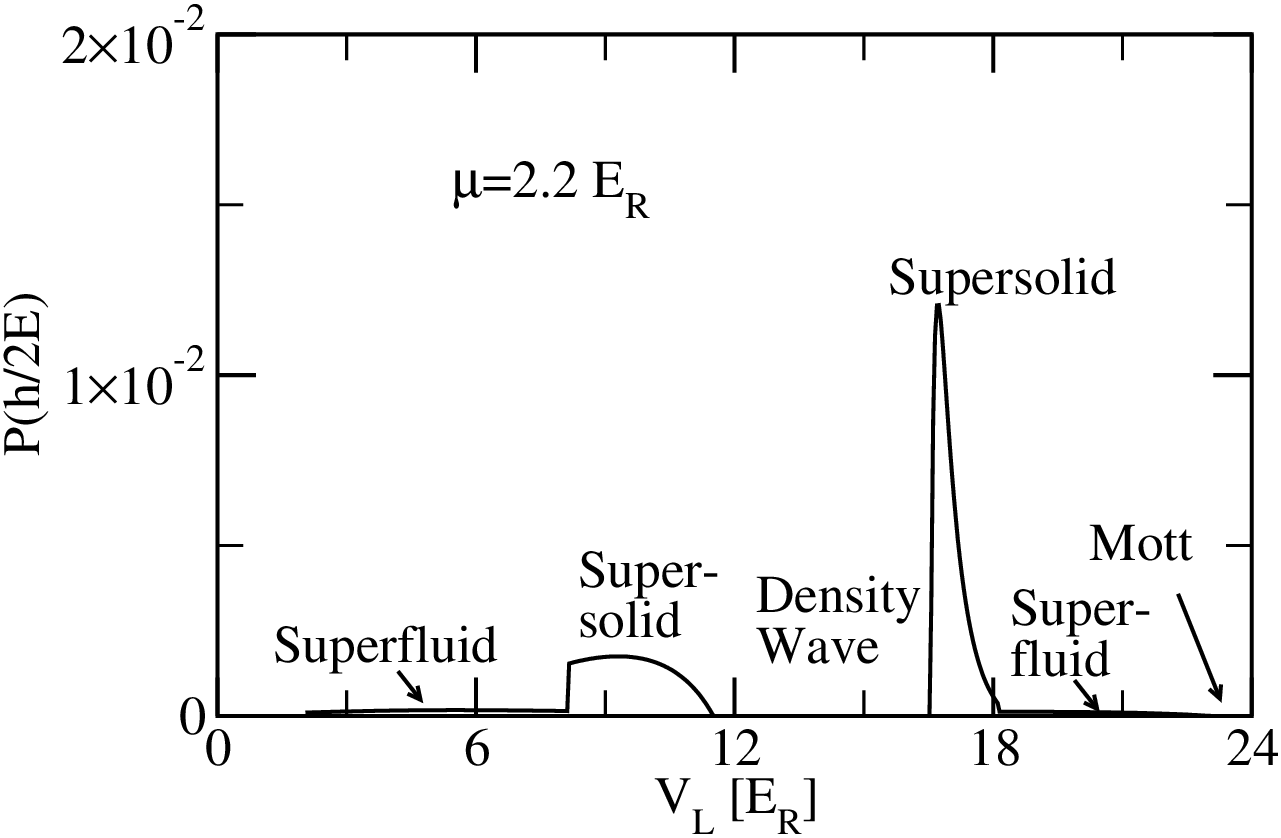}
\caption{ 
The maximum probability that two 
particles from the $(2,2)$ band
occupy the same site and scatter to the $(3,1)$ and $(1,3)$ bands
versus lattice depth with $\mu=2.2 E_R$ and $\bar{g}_2=50$.  
\label{mix}}
\end{figure}

The probability of 
finding two particles at the same site is small in all
regions of Fig.~\ref{phase}.  Ignoring fluctuations, 
the $\rho=1$ Mott and $\rho=1/2$ density wave
phases have {\em no} double occupancy.  Therefore, the superfluid and 
the supersolid phases remain as the only candidate phases involving
on-site, inter-band scattering processes.  
Fig.~\ref{mix} 
plots the maximum scattering probability at $T=h/2E$  
as a function of lattice depth for a chemical
potential $\mu^{(2,2)}=2.2 E_R$ and $\bar{g}_2=50$.  
The probability of finding two particles in 
nearby bands is less than $2\times10^{-2}$.  This suggests that at
intermediate lattice depths and low
densities the single-band ground states resist on-site 
scattering processes into neighboring bands.  The energy difference,
$E$, remained non-zero for all $V_L$ primarily because of the
strong inter-band, nearest neighbor interaction in the neighboring
bands.  

We have shown that promoting bosons to higher bands of optical
lattices can lead to states beyond the superfluid and Mott
states present in zero-range, Bose-Hubbard models of the lowest band.   We 
argue that the ground states of an extended Bose-Hubbard 
model capture the essential physics of bosonic atoms placed in a 
single, isotropic band with a minimum lying, 
energetically, near the top of the optical lattice.  The resulting 
supersolid and density wave states add to the set of observable phases.

Additional phases may arise outside 
the set of approximations leading to Fig.~\ref{phase}.  Gutzwiller, 
mean-field theory should be an excellent approximation for $d>1$ and
affirms results obtained from quantum Monte Carlo studies for
$d=2$ \cite{Wagenblast}.  However, our results overestimate the
strength of the supersolid phase because we have excluded a competing 
phase-separated state \cite{Batrouni}.   Furthermore, strong 
inter-band mixing can populate 
anisotropic neighboring bands ({\em e.g.} (3,1) and (1,3)) leading to 
stripe-like superfluid states which coexist with the superfluid and
supersolid states in the primary band.

The states confined to band 
$\bm{\alpha}_p$ are, technically, metastable.  We require 
$\tau\gg h/\vert t^{\bm{\alpha}_p}_{ij} \vert$,
where $\tau$ is the lifetime of the state.  $\tau$ 
may be affected by dissipative effects including 
collective mode inter-band scattering.   
Our results for the realizable extended Bose-Hubbard Hamiltonian, in
addition to providing a rich quantum phase diagram, yields 
an interesting connection to topological
quantum computation \cite{Freedman} in cold atom optical lattices.

We thank J.K. Jain, K. Park, S. Rolston and Y. Zhang for valuable
discussions.  This work is supported by ARDA, ARO and NSA-LPS.

%%%%%%%%%%%%%%%%%%%%%%%%%%%%%%%%%%%%%%%%%%%%%%%%%%%%%%%%%%%%%%%%%%%%%%%

%%%%%%%%%%%%%%%%%%%%%%%%%%%%%%%%%%%%%%%%%%%%%%%%%%%%%%%%%%%%%%%%%%%%%%%


\begin{thebibliography}{}
\bibitem{inter} We estimate that nearest neighbor, inter-band
processes contribute less than on-site processes.
\end{thebibliography}
\end{document}